\documentclass[]{interact}
\usepackage{mathtools}
\usepackage{epstopdf}
\usepackage{subfigure}
\usepackage{natbib}
\bibpunct[ ,]{(}{)}{;}{a}{}{,}

\renewcommand\bibfont{\fontsize{10}{12}\selectfont}

\usepackage[english]{babel}
\usepackage[utf8x]{inputenc}
\usepackage{algorithm}
\usepackage{algorithmic}
\usepackage[colorlinks=true,
            linkcolor=red,
            urlcolor=blue,
            citecolor=blue]{hyperref}
\usepackage{multirow}
\usepackage{multicol}
\usepackage{lscape}
\usepackage{rotating}
\usepackage{soul}
\usepackage[normalem]{ulem}
\usepackage{slashbox}
\usepackage{footmisc}

\theoremstyle{plain}
\newtheorem{theorem}{Theorem}[section]

\theoremstyle{definition}
\newtheorem{definition}[theorem]{Definition}

\theoremstyle{remark}

\theoremstyle{finding}
\newtheorem{finding}{Finding}

\theoremstyle{observation}
\newtheorem{observation}{Observation}

\theoremstyle{conjecture}
\newtheorem{conjecture}{Conjecture}

\begin{document}



\title{Externalities in {Endogenous Sharing Economy} Networks}
\author{
\name{Pramod~C.~Mane\textsuperscript{a}, Kapil~Ahuja\textsuperscript{a}, and Nagarajan Krishnamurthy\textsuperscript{b}}
\affil{\textsuperscript{a}IIT Indore \\
\textsuperscript{b}IIM Indore}
}
\maketitle

\begin{abstract}
This paper investigates the impact of link formation between a pair of agents on {the} resource availability of other agents {(that is, externalities)} in a social cloud network, {a special case of endogenous sharing economy networks}. Specifically, we study how {the} closeness between agents and the network size affect externalities. 

We conjecture, and experimentally support, that for an agent to experience positive externalities,  an increase in its closeness is necessary. The condition is not sufficient though. We, then, show that for populated ring networks, one or more agents experience positive externalities due to an increase in the closeness of agents. Further, the initial distance between agents forming a link has a direct bearing on the number of beneficiaries, and the number of beneficiaries is always less than that of non-beneficiaries. 
\end{abstract}

\begin{keywords}
endogenous; sharing economy; network; externalities; social cloud
\end{keywords}

\section{Introduction}\label{sec:introduction}
{The sharing economy \citep{thomas-2016-sharing} is not free from externalities, the effects of actions of individual agents on other agents. Studies (for example, \cite{Wenjun-2018-sharing}) have investigated externalities and its sources in the sharing economy. However, these studies have mainly focused on platform-mediated sharing models where consumers and providers are matched. Numerous studies have looked at sharing platforms with negative externalities \citep{vanessa-2015-sharing, Molly-2017-sharing, Koen-2107-sharing} and with positive externalities too \citep{Basselier-2018-sharing}.}

{Externalities in endogenous network formation are usually studied in determining which network structure is likely to emerge, and to understand the tension between network stability and its efficiency \citep{Buechel-over-connected}.} {However, in this paper, we study externalities with a different motivation --- to study how closeness and network size affect externalities.}

{To study the above aspects, we consider the social cloud model proposed by  \cite{social-cloud-network}, a sharing economy network \citep{Leonidasand-2017-sharing}, different from platform-mediated sharing models and endogenously formed. On the one hand, we have online data backup services like BuddyBackup\footnote{\label{footnote:buddybackup}buddybackup.com} and CrashPlan\footnote{\label{footnote:crashplan}crashplan.com}, which are examples of such social clouds, where two agents share or trade resources directly with each other, without an intermediary third-party. On the other hand, we have traditional cloud service providers like Amazon's AWS S3\footnote{aws.amazon.com/s3} and Microsoft's Azure\footnote{azure.microsoft.com}, which are examples of players in a horizontal market where different cloud providers compete for customer requests of computational resources. 
}

\section{{Social Cloud Model}}\label{sec:model}
{A social cloud $\mathfrak{g}$ is a storage or computational resource sharing network of $N$ agents connected by $\mathcal{L}$ links, that evolves endogenously when agents build their resource sharing connections. 
According to \cite{Chard-social-cloud-model}, agents could limit resource sharing with friends who are close to them. \cite{social-cloud-network} capture this closeness by using the harmonic centrality measure \citep{boldi-harmonic-axioms}, as }
\begin{equation*}\label{eq:centrality}
\Phi_i(\mathfrak{g})=\sum\limits_{j \in \mathfrak{g} \setminus \{i\}} \frac{1}{d_{ij}(\mathfrak{g})},
\end{equation*}
where $d_{ij}(\mathfrak{g})$ is the distance between agents $i$ and $j$, that is, the length of any shortest path between them. \\

{Similar to \cite{social-cloud-network}, we define} the probability that agent $i$ obtains a resource from $j$ as
\begin{equation*}
\alpha_{ij}(\mathfrak{g})=\frac{\frac{1}{d_{ij}(\mathfrak{g})}}{\Phi_{j}(\mathfrak{g})}. 
\end{equation*}
\\ 
{The probability that agent $i$ will get the resource from at least one agent in the network $\mathfrak{g}$ is}
\begin{equation*}\label{eq:probnetwork}
\gamma_{i}(\mathfrak{g})=1-\prod\limits_{j\in\mathfrak{g}}(1-\alpha_{ij}(\mathfrak{g})).
\end{equation*}
Based on the definition of externalities by \cite{Jackson}, we define the following for the social cloud model. 
\begin{definition}\label{def:externalities}
Suppose $\mathfrak{g}$ is a social cloud, with distinct agents $i, j, k$ such that $\langle jk \rangle \notin \mathfrak{g}$. If $j$ and $k$ add a link, the resulting network denoted by $\mathfrak{g}+\langle jk \rangle$, then $i$ experiences
				\begin{enumerate}
					\item no externalities if $\gamma_{i}(\mathfrak{g} +\langle jk \rangle)=\gamma_{i}(\mathfrak{g})$,
					\item negative externalities if $\gamma_{i}(\mathfrak{g}+ \langle jk \rangle) < \gamma_{i}(\mathfrak{g})$, and
					\item positive externalities if $\gamma_{i}(\mathfrak{g}+ \langle jk\rangle) > \gamma_{i}(\mathfrak{g})$.
				\end{enumerate}
\end{definition}
{Our objective is to investigate the impact of a newly added link $\langle jk\rangle$ in $\mathfrak{g}$ on the relation between changes in}
\begin{enumerate}
\item {$\Phi_i$ and $\gamma_{i}$,}
\item {$N$, $d_{jk}$ and $\gamma_{i}$, } 
\end{enumerate}
{for all $i\in\mathfrak{g}$.} 
\section{{Findings}}\label{sec:discussion}
\subsection{{Externalities and Closeness}}\label{subsec:closeness}
\begin{figure*}[ht!]
\centering
\subfigure[Network $\mathfrak{g}$ ]
{
\includegraphics[scale=0.15]{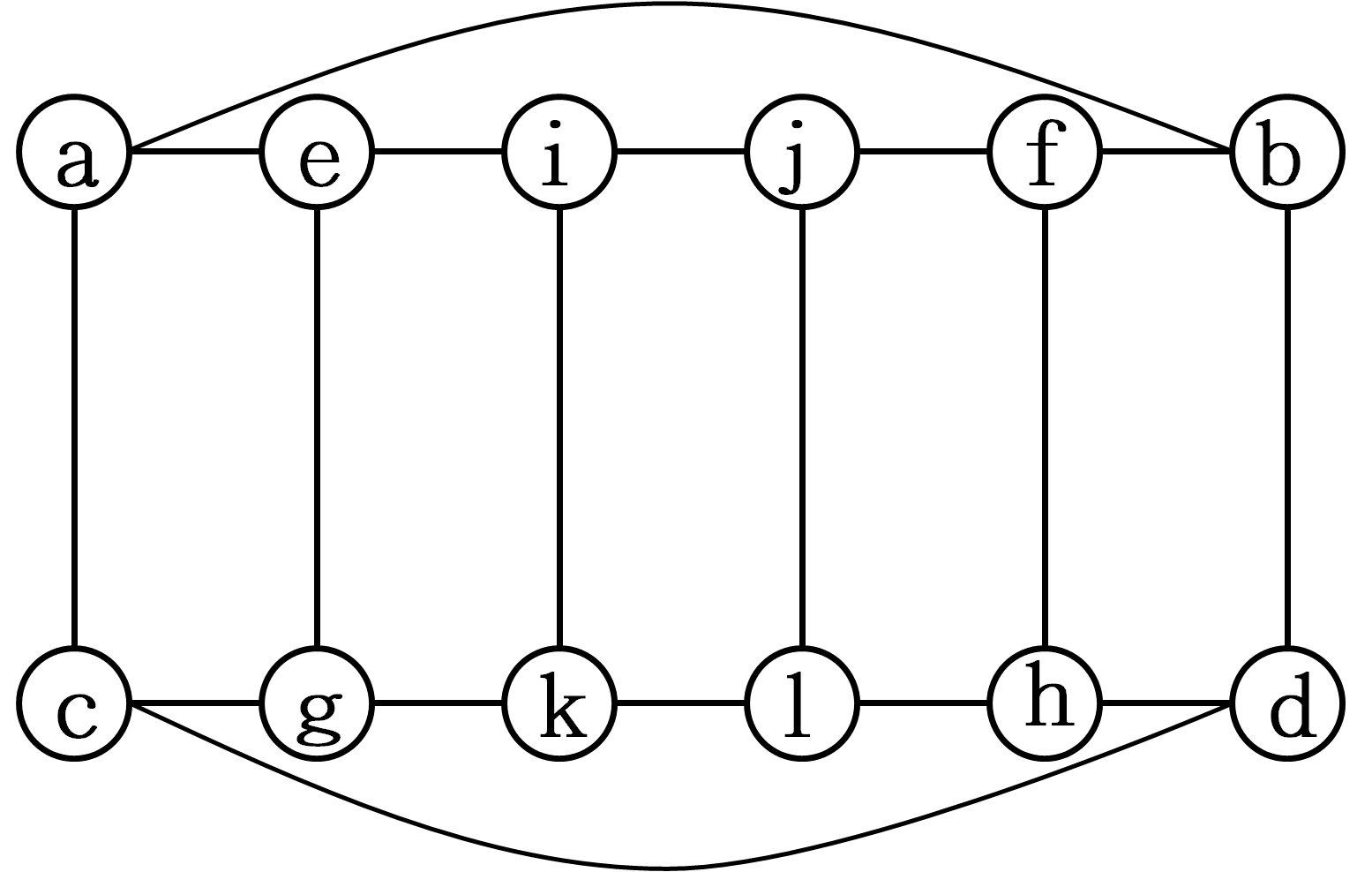} 
\label{fig:externalities-g-network}
}
\quad \quad \quad
\subfigure[Network $\mathfrak{g}+\langle jk\rangle$]
{
\includegraphics[scale=0.15]{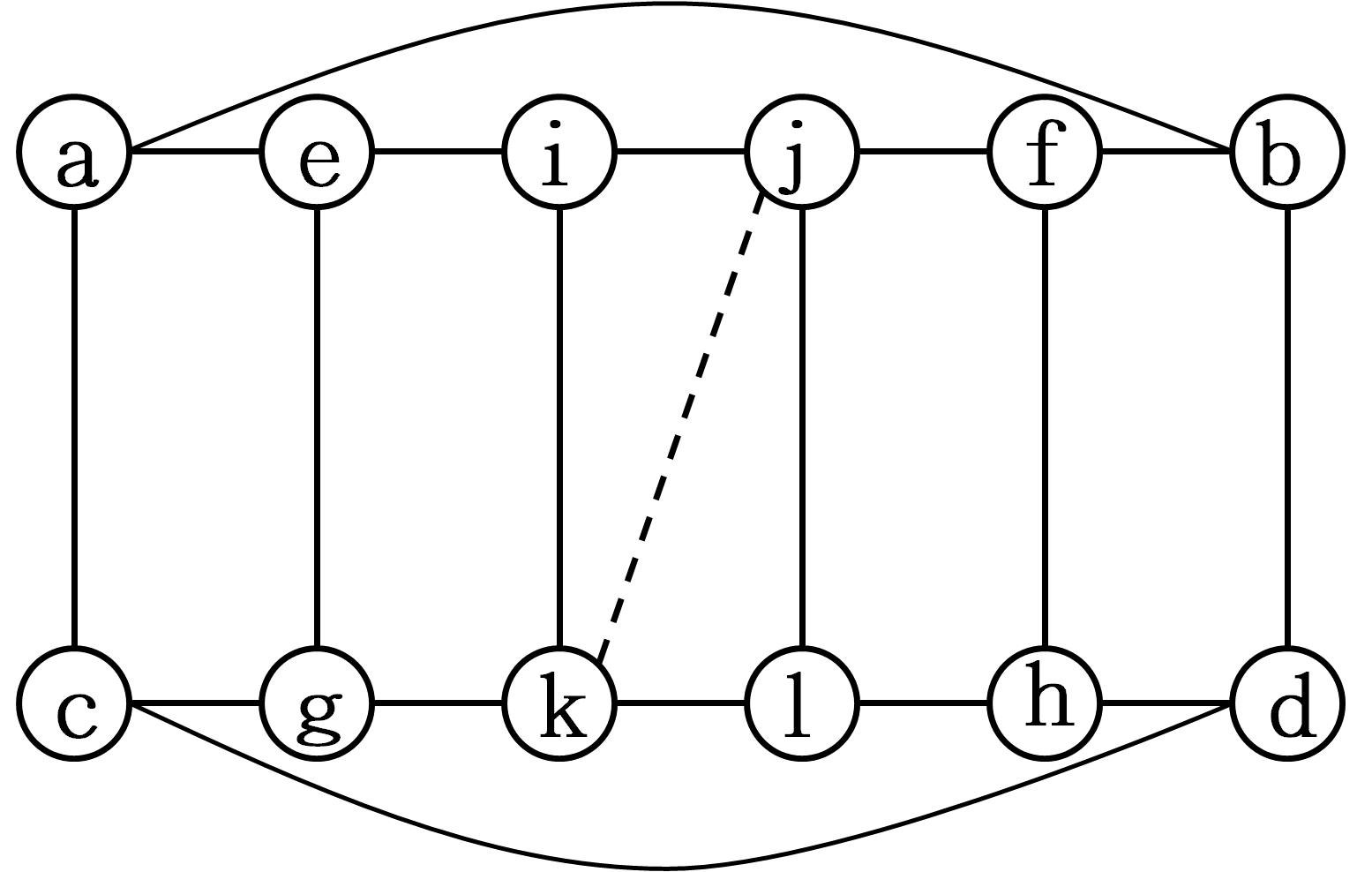} 
\label{fig:externalities-g-bj-network}
}
\caption{Example: Externalities and Closeness}
\label{fig:externalities-first-example}
\end{figure*}

{On first glance, 
it seems that the harmonic centrality would lead to only negative externalities. However, a careful analysis reveals that both positive and negative externalities are exhibited. This is an important motivation of our study.} Consider the network $\mathfrak{g}$ (Figure \ref{fig:externalities-g-network}). On adding link $\langle jk\rangle$, we get $\mathfrak{g}+ \langle jk\rangle$ (Figure \ref{fig:externalities-g-bj-network}).
$\gamma_i(\mathfrak{g}+ \langle jk\rangle)-\gamma_i(\mathfrak{g})$ provides a basis for the study of externalities. As shown in Table \ref{table:probability}, the link $\langle jk\rangle$ is advantageous for agents $f$ and $g$ (shown in blue), and for $j$ and $k$ themselves, but disadvantageous for the remaining agents (shown in red). 

\begin{table}[ht!]
\begin{tabular}{|c|c|c|c|c|c|}
\hline
agent ($x$) & $\Phi_{x}(\mathfrak{g})$ & $\gamma_{x}(\mathfrak{g})$ & $\Phi_{x}(\mathfrak{g}+\langle jk \rangle)$ & $\gamma_{x}(\mathfrak{g}+\langle jk\rangle)$ & $\gamma_{x}(\mathfrak{g}+\langle jk\rangle)-\gamma_{x}(\mathfrak{g})$ \\ \hline
$a$         & 6.25                     & 0.654                      & 6.25                                        & 0.646                                        & \textcolor{red}{-0.008}                                                                                                           \\ \hline
$b$         & 6.25                     & 0.654                      & 6.33                                        & 0.650                                        & \textcolor{red}{-0.004}                                                                                                           \\ \hline
$c$         & 6.25                     & 0.654                      & 6.33                                        & 0.650                                        & \textcolor{red}{-0.004}                                                                         \\ \hline
$d$         & 6.25                     & 0.654                      & 6.25                                        & 0.646                                        & \textcolor{red}{-0.008}                                                                                                           \\ \hline
$e$         & 6.25                     & 0.654                      & 6.25                                        & 0.644                                        & \textcolor{red}{-0.011}                                                                                                           \\ \hline
$f$         & 6.25                     & 0.654                      & 6.50                                        & 0.657                                        & \textcolor{blue}{0.003}                                                                                                            \\ \hline
$g$         & 6.25                     & 0.654                      & 6.50                                        & 0.657                                        & \textcolor{blue}{0.003}                                                                                                            \\ \hline
$h$         & 6.25                     & 0.654                      & 6.25                                        & 0.644                                        & \textcolor{red}{-0.011}                                                                                                           \\ \hline
$i$         & 6.25                     & 0.654                      & 6.25                                        & 0.637                                        & \textcolor{red}{-0.017}                                                                                                           \\ \hline
$j$         & 6.25                     & 0.654                      & 7.00                                        & 0.687                                        & 0.033                                                                                                            \\ \hline
$k$         & 6.25                     & 0.654                      & 7.00                                        & 0.687                                        & 0.033                                                                                                            \\ \hline
$l$         & 6.25                     & 0.654                      & 6.25                                        & 0.637                                        & \textcolor{red}{-0.017}                                                                       \\ \hline
\end{tabular}
\caption{Externalities and Closeness}
\label{table:probability}
\end{table}

{\begin{observation}\label{rmk:conditions-remark}
Addition of a link ($\langle jk\rangle$, above) results in the same closeness for some agents (for example, $h$ and $i$) and an increased closeness for the others (for example, $b$ and $c$). Addition of a link cannot decrease the closeness of any agent, and always increases the closeness of the agents who add the link ($j$ and $k$).  
\end{observation}
\begin{conjecture}\label{remark:sufficiency} 
For an agent to experience positive externalities, an increment in its closeness is necessary.  
\end{conjecture}
\begin{finding}\label{remark:not-sufficient} 
For an agent to experience positive externalities, an increment in its closeness is not a sufficient condition. For example, although the closeness of agents $b$ and $c$ increases, their chance of obtaining a resource does not improve.
\end{finding}
}
{
\subsection{Network Size Analysis}\label{subsec:Experiments}
For our second objective, we focus on the ring network with sizes varying from $4$ to $30$ agents, considered reasonably large in the literature on experimental research on  economic networks \citep{Choi-Survey}. Very large networks may encode a very limited amount of information \citep{Chandrasekhar-2016-Experiments}.}

{We focus on the ring network as its harmonic centrality is uniform. This helps us investigate the relation between externalities and network size. }

{To obtain data, we adopt a single computer program-based simulation \citep{naylor1967computer, friedman2004economists} method due to unavailability of data of real world networks like BuddyBackup or CrashPlan, and because of the endogeneity issue \citep{Choi-Survey} faced by various studies.}

We, initially, compute $\gamma_{i}(\mathfrak{g})$ for all agents in a ring network $\mathfrak{g}$. Then, we select an agent $j$ arbitrarily and add a link with another agent $k$ whose distance is two hops from $j$. We compute $\gamma_{i}(\mathfrak{g}+\langle jk \rangle)-\gamma_{i}(\mathfrak{g})$, 
{and count the number of beneficiaries (NOB) $i$ for whom this difference is positive.} 
We repeat the above for all agents {$k$} who are located at distance two hops from agent $j$ in $\mathfrak{g}$. Then, we increment the distance {between $j$ and $k$} by one and follow the same procedure. 
We do this until we exhaust all agents {$j$}. 

\begin{figure*}[ht!]
\centering
\subfigure[Network Size 4-10]
{
	\includegraphics[scale=0.23]{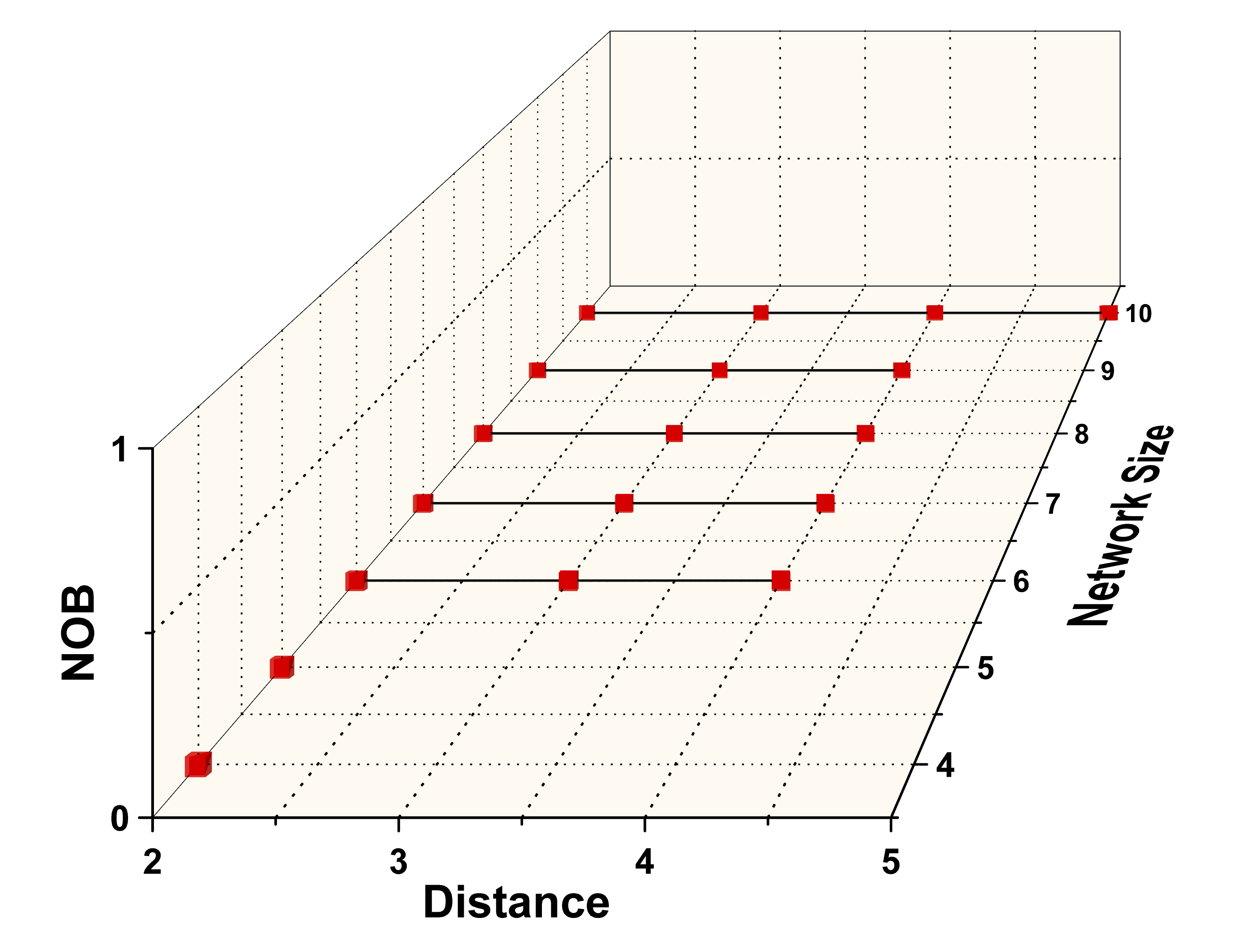} 
	\label{fig:ns-4-9}
}
\quad
\subfigure[Network Size 11-20]
{
	\includegraphics[scale=0.23]{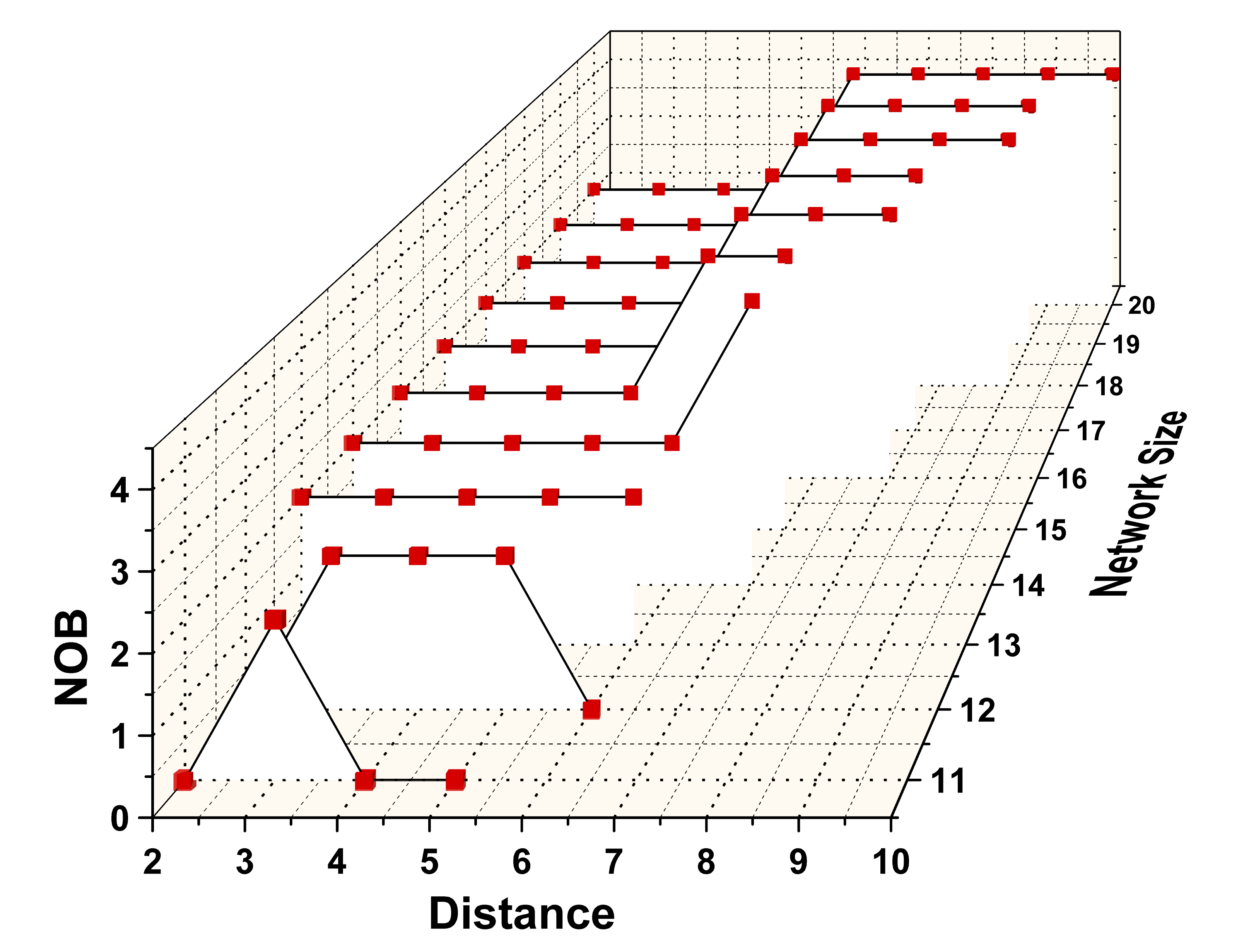} 
	\label{fig:ns-10-20}
}
\quad
\subfigure[Network Size 21-30]
{
	\includegraphics[scale=0.23]{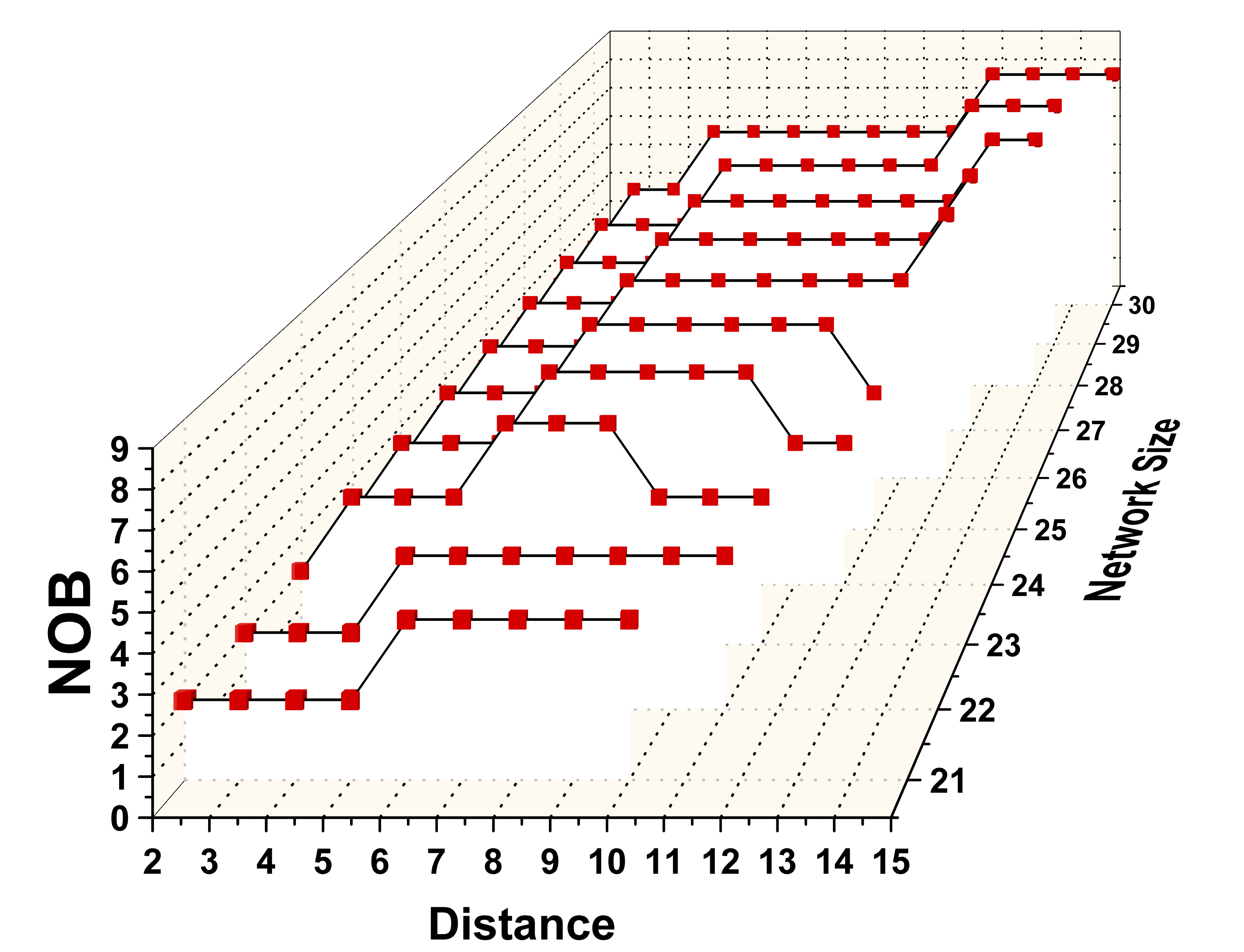} 
	\label{fig:ns-21-30}
}
\quad
\subfigure[Network Size 22-24]
{
	\includegraphics[scale=0.23]{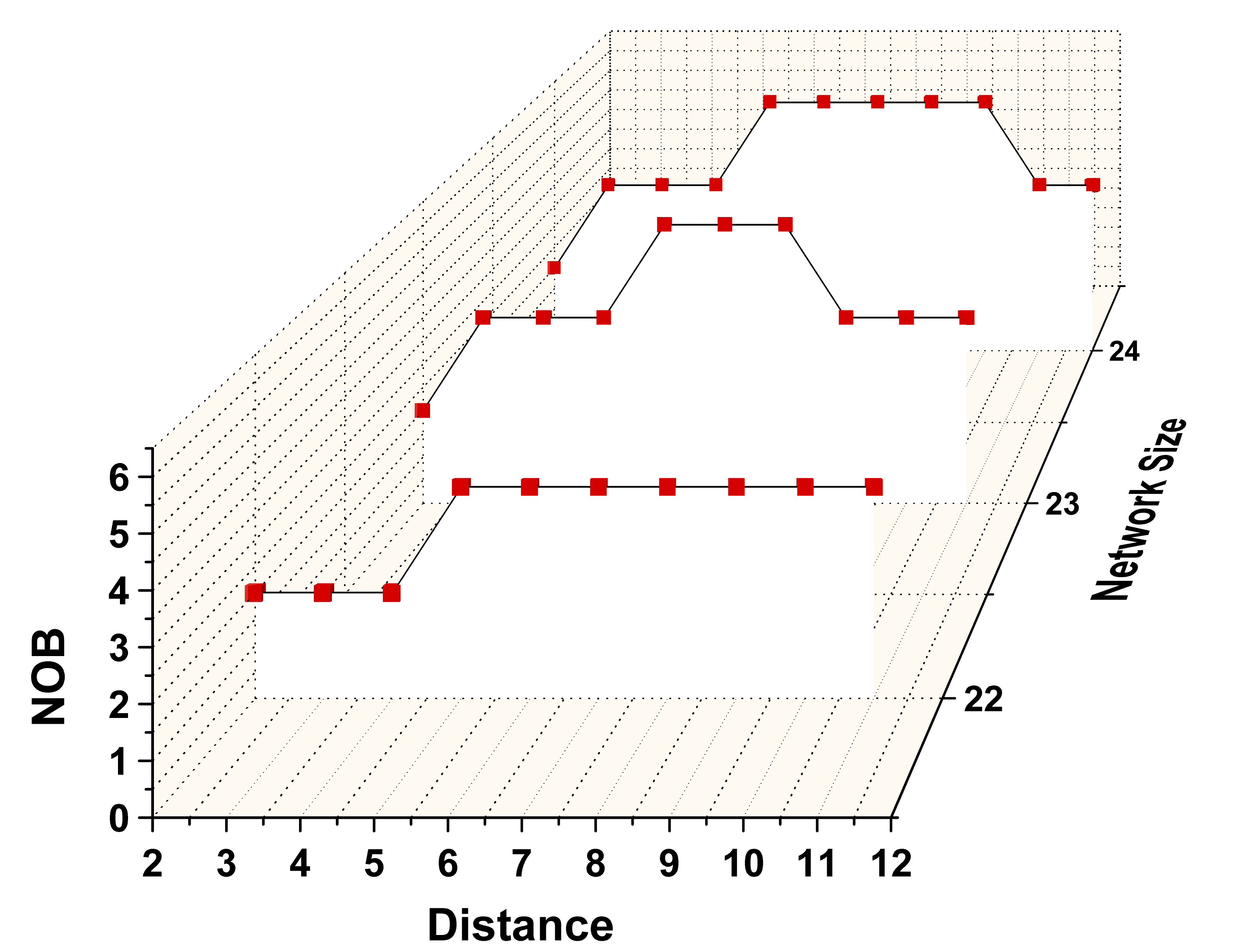} 
	\label{fig:ns-22-24}
}
\caption{Externalities and Network-Size}
\label{fig:network-size-analysis}
\end{figure*}

Figure \ref{fig:network-size-analysis} summarises our results. The $x$-$axis$ represents the shortest distance between two agents involved in  link formation. The $y$-$axis$ represents the NOB. The $z$-$axis$ represents the network size (the number of agents in the network). 

\begin{finding}
In ``less" populated ring networks, no agent experiences positive externalities.
\end{finding} 
In networks with size varying from 4 to 10 (Figure \ref{fig:ns-4-9}), no agent experiences positive externalities. From Conjecture \ref{remark:sufficiency}, positive externalities require an increase in closeness, which is absent here. However, from network size 11 to 30 ( Figures \ref{fig:ns-10-20} and \ref{fig:ns-21-30}), a significant number of agents experience positive externalities. 

\begin{finding}
In ring networks of size greater than 10, as the distance between the agents involved in link addition increases, the number of beneficiaries increases in most cases, and in all cases for small distances. 
\end{finding}
Plots in Figures \ref{fig:ns-10-20}, \ref{fig:ns-21-30} and \ref{fig:ns-22-24} are of this type. {It is clear that, if an agent experiences positive externalities, its closeness should increased (from Conjecture \ref{remark:sufficiency}). This is intuitive --- if a  pair of agents who are far from each other in $\mathfrak{g}$ form a link then, this link reduces the mutual distances among other agents, and therefore, their closeness increase. As discussed in Finding \ref{remark:not-sufficient}, an increase in closeness is not a sufficient condition for positive externalities. We believe that an increase in closeness together with some conditions will always imply positive externalities, and leave open this question of finding the other conditions of sufficiency. }

\begin{finding}
In ring networks, the number of beneficiaries is always less than the number of non-beneficiaries.
\end{finding} 
In all our experiments, the percentage of beneficiaries varies from $0\%$ to $26\%$ of the total number of agents in the network.

\section{Conclusion}\label{sec:conclusion}
{Though our investigations are on a small scale, our findings have many applications and implications.}
{Our study enhances our understanding of externalities in the sharing economy network model. 
Our findings show that network size is also a key factor in determining externalities.}

{Our approach of relating externalities and group well-being (that is, measuring the percentage of beneficiaries versus non-beneficiaries) enriches the transfer-based network formation model (where agents subsidise others to form or not to form a link with others),  \citep{BLOCH200783} by incorporating group subsidisation. In particular, agents can subsidize a pair of agents involved in a link formation, instead of individual subsidization. For  example,  in  some  research  and  development  settings,  where  a  set of beneficiary-firms are willing to pay a pair of firms that would like to collaborate \citep{Jackson-Book}.  Using the approach and results discussed in the paper, one can model this situation either as collective subsidization or as bargaining on link formation.  In this case, a set of non-beneficiary-firms will pay for the pair of firms to not collaborate with each other, or a set  of beneficiary-firms will pay for the pair of firms to collaborate with each other.  Alternatively, both beneficiaries and non-beneficiaries may bargain for link formation.}

{Our result regarding the network size and the distance between agents can be used to formulate resource allocation policy in social clouds. Specifically, the policy may include the use (or recommendation) of friends as backup partners, so as to avoid negative externalities or minimize the number of non-beneficiaries (for example, by choosing agents for data backup who are far away from the agent requesting backup).  Although our results are for ring networks, they can be extended to other network structures too.}

{
Existing literature has not looked at resource availability in social networks where links (connections) are endogenously formed. We believe that our study is a first step towards enhancing our understanding of externalities in social clouds, and hence in endogenous sharing economy networks.
}


\end{document}